\documentclass[conference]{IEEEtran}
\IEEEoverridecommandlockouts
\usepackage{cite}
\usepackage{amsmath,amssymb,amsfonts}
\usepackage{algorithmic}
\usepackage{algorithm}
\usepackage{graphicx}
\usepackage{textcomp}
\usepackage{xcolor}
\usepackage{algorithm}
\usepackage{tikz}
\usepackage{url}
\usetikzlibrary{positioning}

\def\BibTeX{{\rm B\kern-.05em{\sc i\kern-.025em b}\kern-.08em
    T\kern-.1667em\lower.7ex\hbox{E}\kern-.125emX}}
\begin{document}

\title{AI/ML Model Cards in Edge AI Cyberinfrastructure: towards Agentic AI\\
\thanks{This research funded in part through grants from the National Science Foundation OAC-2209872, OAC-2112606 and reflects the views of the authors alone. Corresponding author Beth Plale, plale@uoregon.edu}
}

\author{
  \IEEEauthorblockN{Beth Plale}
  \IEEEauthorblockA{\textit{School of Computer and Data Sciences}\\
                    \textit{University of Oregon}\\
                    Eugene, OR USA\\
                    0000-0003-2164-8132}
\and
                    
                    \IEEEauthorblockN{Neelesh Karthikeyan}
  \IEEEauthorblockA{\textit{School of Computer and Data Sciences}\\
                    \textit{University of Oregon}\\
                    Eugene, OR USA\\
                    0009-0009-9817-7042}
  \and
  \IEEEauthorblockN{Isuru Gamage}
  \IEEEauthorblockA{\textit{Intelligent Systems Engineering}\\
                    \textit{Indiana University}\\
                    Bloomington, IN USA\\
                    agamage@iu.edu}
  \and
    \IEEEauthorblockN{Joe Stubbs}
    \IEEEauthorblockA{\textit{Texas Advanced Computing Center (TACC)}\\
                    \textit{University of Texas}\\
                    Austin, TX USA\\
                    jstubbs@tacc.utexas.edu}
\and
  \IEEEauthorblockN{Sachith Withana}
    \IEEEauthorblockA{\textit{School of Computer and Data Sciences}\\
    University of Oregon\\
                     Eugene, OR USA\\
                    0000-0002-2022-8155}

}

\maketitle

\begin{abstract}
 AI/ML model cards can contain a benchmarked evaluation of an AI/ML model against intended use but a one time assessment during model training does not get at how and where a model is actually used over its lifetime. Through Patra Model Cards embedded in the  ICICLE AI Institute software ecosystem we study model cards as dynamic objects. The study reported here assesses the benefits and tradeoffs of adopting the Model Context Protocol (MCP) as an interface to the Patra Model Card server. Quantitative assessment shows the overhead of MCP as compared to a REST interface.  The core question however is of active sessions enabled by MCP; this is a qualitative question of fit and use in the context of dynamic model cards that we address as well.

%This early result is part of an ongoing exploration of the benefit that Model Context Protocol (MCP) can bring to the AI accountability target of the Patra Model Cards framework. 
\end{abstract}


\section{Introduction}

AI/ML models are widely used in scientific research.  Their traceability (where and in what manner used, how one AI/ML model is related to another, etc) is often overlooked. But for responsible use of AI and to increase the reproducibility of science, this traceability is important. AI/ML model cards are a step in the right direction. 

 An AI/ML \textit{Model Card} is a structured overview of how an AI/ML model was designed and evaluated. Mitchell et al. describe the original intent behind model cards this way: "model cards are short documents accompanying trained machine learning models that provide benchmarked evaluation in a variety of conditions, such as across different cultural, demographic, or phenotypic subgroups that are relevant to the intended application domains"~\cite{mitchell2019model}. Model cards further are used to disclose the context under which models are intended to be used, details of the performance evaluation procedures, and other relevant information. As Mitchel et al. state, model cards provide benchmarked evaluation that is necessary for the informed use of a model. 
 
 Model Cards, which have seen wide adoption since their introduction, are an entry into strengthening \textit{responsible AI practice} which itself includes \textit{explainability} which provides humans with clear, understandable, and meaningful explanations for why an AI system made a certain decision, \textit{interpretability}, which focuses on understanding the inner workings of how a model’s inputs influence its outputs, and \textit{accountability} which gets at who is accountable for the AI systems when something goes wrong.  Big tech companies like Amazon, Hugging Face, and Google are prominent in their support for model cards and provide tools for card creation and persistence. For instance, the  Amazon SageMaker has its Model Card that is integrated in the SageMaker Model Registry.  With the broad uptake of Large Language Models, the debate around responsible AI has intensified.  And some tech providers are responding. For instance, in summer of 2023 Meta and Microsoft announced the launch of Llama2.  The cited research paper behind the release is extensive in its examination of the model through the lense of responsible AI~\cite{touvron2023llama}.  The paper includes its Model Card in the appendix (though the Model Card hews to the original Mitchell et al intent and uses the research paper itself for greater exposition.)

We acknowledge and celebrate all forms of AI/ML model cards as a step in the right direction. \textit{Our broader vision of the utility of AI/ML model cards focuses on whether benchmarked evaluation of an AI/ML model can be treated not as a one-time thing, done during model training, but instead can be used
to assess and evolve one’s understanding of a model’s fit
for purpose during its lifetime of use.} Can we particularly collect information that might point to an ill suited use? This problem has motivated us to develop the Patra Model Card framework \cite{withana2024patra}. We view an AI/ML model card not just as manual documentation created at time of model creation, but as an active reflection of model usage through time which can, thus, if practicable, make greater contributions to responsible AI. Our study is focused on AI/ML inference at the edge, and is carried out in the context of the evolving ICICLE AI Institute computational environment~\cite{Panda2024AIMag}.

Edge environments present unique challenges (resource constraints, workload variability, and discrete changes in quality requirements) that static model deployment strategies cannot adequately address. For instance, camera traps in wildlife monitoring may experience sudden bursts when animals appear, requiring rapid adaptation between accuracy-optimized and latency-optimized models. Traditional edge AI deployments rely on fixed model choices that either optimize for performance or require manual reconfiguration when conditions change. This approach fails to capture the dynamic nature of edge workloads, where motion detection, lighting changes, or seasonal patterns can dramatically shift the optimal balance between inference accuracy and response time.

 We have gained experience with model cards through Patra's use in the NSF AI Intelligent CI with Computational Learning in the Environment (ICICLE) Institute ~\cite{Panda2024AIMag} which is developing novel AI-driven pipelines for scientific research at the edge. This is through, in part, by working with MLFieldPlanner, a configurable cyberinfrastructure that can utilize the full edge-cloud-HPC continuum, analyze ML pipelines, and study edge-to-center tradeoffs in model placement. Researchers use ML Field Planner to ``configure experiments to run on real IoT hardware, configure machine learning models to analyze custom benchmark datasets, and experiment with different algorithm configurations, such as storage compression, all from a graphical user interface"~\cite{StubbsMLFieldPlanner}.
 
The motivating use case for our work is AI/ML model deployment at the edge in support of field research~\cite{StubbsMLFieldPlanner}. The scenario is of a researcher deploying AI-enabled devices for multi-modal observation in a remote environmental setting. For example, an ecologist may evaluate the compatibility, latency, and recall of different versions of a YOLO object detection model~\cite{redmon2016lookonceunifiedrealtime} deployed to various edge devices, such as Raspberry Pis and the NVIDIA Jetson family, each equipped with a motion-activated camera. Patra Model Cards coordinate and interoperate with the TapisUI workflow platform~\cite{StubbsMLFieldPlanner} to facilitate edge-based, plug-and-play model orchestration. Through TapisUI, a scientist queries Patra  for models that fit their needs (for example, models optimized for less than 100 ms edge inference latency). Relevant candidate models are directly retrieved from their distributed hosting platforms such as HuggingFace or GitHub through persistent IDs maintained in the model card.  

As we have accumulated experience with use of model cards in ICICLE, and accumulated the model cards themselves, we undertook to explore how the Patra Model Card repository can beneficially serve in an agentic AI setting.  An Agentic AI system is an AI system that exhibits adaptability in achieving complex goals in complex environments with limited direct supervision~\cite{governingAgentic}. We particularly study the tradeoffs of supporting the Model Context Protocol (MCP)\cite{MCP}, an interoperability protocol for connecting AI applications to external systems. MCP offers itself as a standardized way for AI applications to connect to external systems for "connecting AI assistants to the systems where data lives, including content repositories, business tools, and development environments" \cite{MCP}.

We make the following observations about the Patra framework as we consider the viability of MCP as a replacement or additional interface to the REST API.  These observations guide and constrain next steps:
\begin{enumerate}   
    \item When models are frequently deployed their model cards are subject to continuous streaming ingest of information about model use in the ICICLE environment. A model card than thus become quite large (in the GB range.)   
    \item  Patra uses a graph database (currently Neo4j and Cypher queries) that is fronted with a rich API implemented as Flask endpoints. Graph databases have rudimentary schemas, that is, of edges and vertices (nodes).  Patra layers meaning on top of this rudimentary schema through 1) ascribing properties to nodes (and making heavy use of these properties) and 2) through embedding pathways and delineations of subgraphs by logic in the endpoints.  %   \\ column in table could be "can layer supports endpoint logic to validate, delineate, form queries, etc.
    %    \item integrity of the data could be compromised by uncontrolled access 
\end{enumerate}

In this article and contextualized within the larger setting of the ICICLE Institute software ecosystem, we make the following contributions:  
\begin{itemize}
\item foundational principles behind Model Cards as dynamic objects 
\item use case of model cards as dynamic objects in their integration and use in the ICICLE ecosystem
    \item quantitative results of two MCP implementations in comparison to the currently supported REST API 
      \item assessment of the benefits and tradeoffs of adopting the Model Context Protocol (MCP) as an interface to the Patra Model Card server.  The core question is of active sessions enabled by MCP; this is a qualitative question of fit and use in the context of dynamic model cards.
\end{itemize}

The remainder of this article identifies the principles underlying the dynamism in model cards in Section ~\ref{sec:method}. Following that is a motivating use case in Section~\ref{sec:use}, a discussion of system design and MCP in Section ~\ref{sec:design}, experimental results in Section~\ref{sec:experim}, and a discussion in Section~\ref{sec:disc}.  Related work and Conclusions round the paper out in Sections~\ref{sec:related} and \ref{sec:concl} respectively.

\section{Methodology \label{sec:method}}

An AI/ML model card should tightly relate to the AI/ML model it describes - always accessible to lend clarity on what the model is, how it has been and will continue to be used. This is a tall order in an environment where AI/ML Models can be arbitrarily downloaded, transformed, and run anywhere. We have the benefit of exploring the utility of model cards by constraining the working environment to the broader ICICLE cyberinfrastructure environment. This execution environment is underpinned by the Tapis hosted API platform~\cite{tapis}. 

In this section we provide the foundation underlying Patra's dynamic model card framework. 
  
\textbf{AI/ML Model Lifecycle} We model edge AI/ML models deployed at the edge as being in one of the following four states (see Figure ~\ref{fig:lifecycle}): 

\begin{figure}[htb]   \centering  \includegraphics[width=7cm]{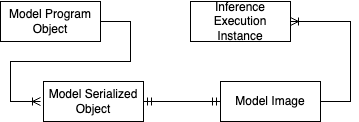}  \caption{AI/ML model object lifecycle}   \label{fig:lifecycle} \end{figure}

\begin{itemize}
\item{\textit{Model Program Object}}. An AI/ML model begins its life in some state of training. Model Program Objects are often Python objects, are subject to training, and reside in a repository such as Github. 
\item{\textit{Model Serialized Object}} has undergone
serialization for sharing, packaging into a container, etc. Serialized models are persisted in an object store, file system, etc. Model serialization can occur any number of times so the n-ary relationship between a model program object and a model serialized object is 1:M.  

\item{\textit{Model Image}} is a serialized model packaged for deployment as an inference server. This packaging makes a serialized model remotely accessible and deployable. A Model Image is frequently maintained as a persistent artifact in an image server such as DockerHub. 
\item{\textit{Inference Execution Instance}} represents an inference server, a running executable. It is a time-bounded instance of model while executing. A single Model Image can be deployed repeatedly, hence a model image and its execution instance are in a 1:M relationship. While all artifacts have properties, the Inference Execution Instance artifact has temporal and spatial properties that capture its time and location of execution as well as such things as pattern of use. 
\end{itemize}

\begin{figure}[htb]
    \centering
    \includegraphics[width=7cm]{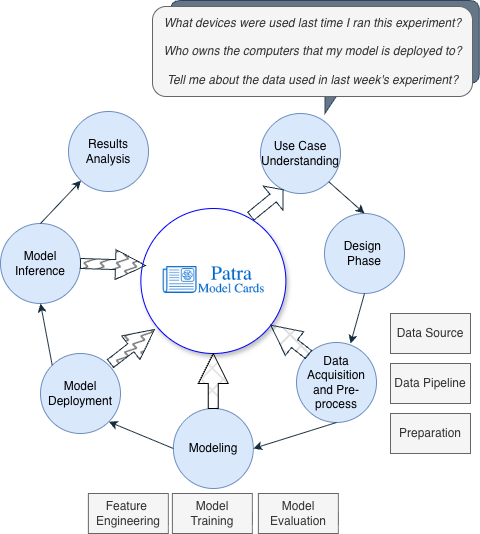}
    \caption{AI/ML model development-deployment pipeline with model card collection points}
    \label{fig:archit}
\end{figure}

\textbf{Model Development/Deployment Pipeline.} 
Patra Model Cards operate  within a development pipeline in the ICICLE AI ecosystem. The pipeline has multiple phases from design through training, deployment, use, and analysis. This is depicted in Figure ~\ref{fig:archit} starting in the upper right hand corner.

\textit{Model Card as Design Aid.} The lifecycle of a deployment pipeline begins with a researcher/educator querying the Patra data source to gain an understanding of AI/ML models that are available for use, for instance by asking questions that get at issues of their prior history and use, such as:  \textit{"What devices were used last time I ran this experiment?}, \textit{"Who owns the computers that my model will be deployed to?"}, and \textit{"Tell me about the data used in last week's experiment}. This activity is shown in the upper right bubble "Use Case Understanding". %The Patra repository database to answering these questions.   

\textit{Model Onboarding.} The cross-hatch arrows pointing inward in the figure are static information collection points. During model training, a researcher interacts through their Jupyter notebook to provide basic information about their model (through the Patra ModelCard toolkit). The toolkit supports auto-population of key metadata, for instance fairness and XAI scanners. The toolkit  validates and outputs the Model Card as a structured JSON file. The actual model file (for example, a PyTorch .pt file) is uploaded to an artifact store such as HuggingFace or GitHub. The model card entry is updated with a link to the artifact’s location. At this point, the model becomes available in the system’s model catalog.

\textit{Runtime Data Capture.} The zig-zag arrows pointing inward in Figure~\ref{fig:archit} are runtime information collection points.  
Patra continuously captures model behavior data through the CKN streaming system~\cite{withana2023ckn}, our earlier work, during runtime inference at the edge. Each edge server runs a CKN daemon that streams events (prediction outcomes, accuracy, latency, CPU/GPU usage, etc.) through a Kafka message broker. These events are consumed by CKN stream processors; Patra subscribes to these streams which are the source of information about execution instances for the Model Card. The daemon captures real-time events from edge devices, augmenting them with details such as model usage, resource consumption, prediction accuracy, and latency. Patra then merges this data into its graph. Each inference execution becomes a node (an “execution instance” of the model) that is connected to the corresponding Model Card. %In sum, real-time deployment data are automatically ingested and linked in Patra’s KG so that the Model Card reflects the model’s actual operational profile over time.

\textit{Automated Model Selection.}
While not directly reflected in Figure~\ref{fig:archit}, when it is a system rather than a human choosing an inference model and because the entire Model Card is machine-actionable, an orchestrator can issue full-text or graph queries against Patra. In practice, a controller might call the /search endpoint with keywords or criteria (for example, “camera-trap classification”) to retrieve matching Model Cards. It can filter results by fairness or explainability annotations embedded in the card. Alternatively, direct graph queries on performance attributes (e.g., selecting the model with the highest accuracy under a latency threshold) can be executed via Patra’s REST interface. 

Patra provides a “machine-actionable API” enabling automated selection based on fairness, explainability, and performance metrics. In other words, the knowledge graph can be queried to rank models by these attributes. Once a suitable model ID is identified, its linkset can be resolved and the model artifact retrieved for deployment. These selection steps are grounded in the provenance recorded in Patra, so as to ensure that choices are traceable. If multiple models satisfy the criteria, Patra’s built-in similarity analysis can even infer versioning relations (e.g., alternateOf or revisionOf) to guide the selection of the most appropriate variant. 
Thus, automated model selection is achieved by programmatically querying the Patra-KG and using the integrated metadata and metrics to inform the decision.

\begin{figure}
    \centering
    \includegraphics[width=8.0cm]{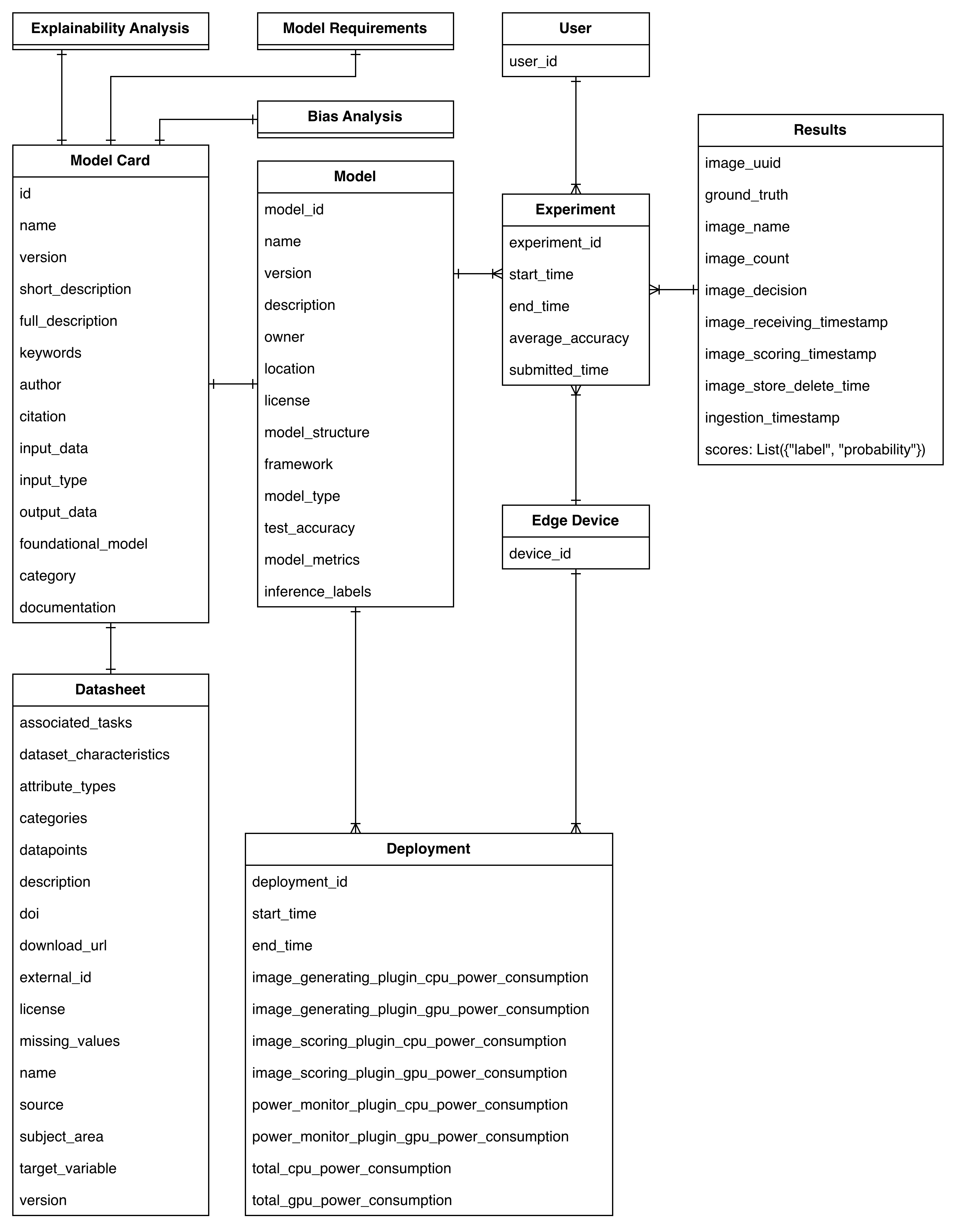}
    \caption{Core Entities of the Patra Model Card}
    \label{fig:dataModel}
\end{figure}

\textbf{Core Entities.} The high level structure of a model card is captured in the E-R diagram in Figure~\ref{fig:dataModel}. Shown are four nodes: Model Card, Model, Edge Server, and Deployment. Each node has properties (what is shown is a subset). Note that a Model Card node and Model node are in a 1:1 n-ary relationship.  The model is described in its own node.  A Model is in a relation with its Deployment.  The relation (edge) is named, though the named edge is not depicted here.  A model can be deployed repeatedly, as is captured by the 1:m n-ary relationship between a model and an intstance of its deployment. A deployed model serves the role of the Inference Execution Instance of Figure~\ref{fig:lifecycle} in that it captures information about the behavior of inference server in use. 

Persistent storage is through a graph database (Neo4j). Graph databases enable constant-time edge traversals and multi-hop lineage queries without expensive join patterns, which is attractive for the type of provenance relationships that are represented in our model cards. Patra's model card search endpoint utilizes Neo4j's full-text index to search for model card nodes by a text prompt and rank results by relevance score. The same query can be replicated in relational database management systems like PostgreSQL and MySQL.

It should be noted that it is a design decision that we make to use graph nodes to represent larger conceptual objects, and turn to properties of nodes for the attributes (details) to describe the nodes.  As we will later see, queries are simpler as a result of this decision.  Retrieving a model card requires a handful of node-level retrievals. But embedded properties poses an additional burden on interpretation as we will discuss in Section~\ref{sec:disc}.
%beth: to get visibility inside a node we need indexed fields.  What decisions have we made about what to index?  Address that here.

\textbf{FAIR Signposting.}
FAIR Signposting~\cite{fairsignposting} is endpoint retrieval of minimal information about objects using just a REST HEAD request which is lightweight especially for large return result sets. An AI/ML model has a unique ID that is frequently a URL pointing to the model in the repository (e.g., HuggingFace). The Patra Model Card unique ID is a unique string made up of author, model name, and version fields. 

The model card ID can be used to navigate the model’s metadata using \textit{Link-Set} endpoints. For instance, a HEAD request to \texttt{/modelcard/<ID>} returns HTTP Link headers that point to the model’s semantic metadata (the “modelcard linkset”), and a GET to \texttt{/modelcard/{id}/linkset} returns the full linkset JSON for that model. These responses follow the Signposting conventions: they expose URLs to the Model Card itself, the container image, datasheets, and any other related artifacts. In effect, the Patra KG acts as a signposting broker: given a model’s PID, one can resolve to all connected resources in the graph via these standard APIs. %The mechanism supports reproducibility and discovery, since clients can programmatically follow links from the PID to retrieve the model’s performance data, fairness metrics, or container image without ambiguity.

 \section{Use Case - Training to Deployment Pipeline \label{sec:use}}

\begin{figure}
    \centering
    \includegraphics[width=8.5cm]{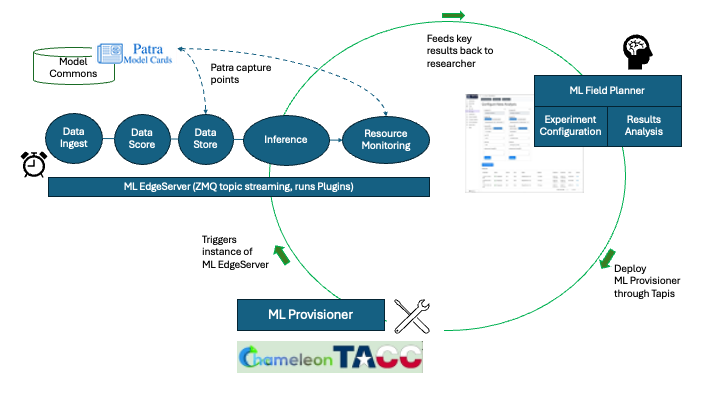}
    \caption{Inference at the Edge through ML Workbench}
    \label{fig:useCase}
\end{figure}

We illustrate AI/ML model cards at the edge through a use case in the ICICLE AI Institute ecosystem.  As was in Figure~\ref{fig:archit} the upper right hand corner is the starting point for the pipeline in Figure~\ref{fig:useCase}.  In the current figure, the researcher configures an experiment to run on the edge using the MLFieldPlanner tool and GUI~\cite{StubbsMLFieldPlanner}. With an experiment configured, an instance of MLProvisioner is triggered as a Tapis job.  

The provisioner reads the experiment configuration and provisions the hardware. Supported hardware includes OpenStack platforms such as Chameleon~\cite{chameleon} and edge devices with dedicated IP addresses.  The configured hardware is brought up running an instance of MLEdgeServer.  MLEdgeServer supports plugins that carry out functionality (including data scoring, inference, resource monitoring). These plugins communicate through ZMQ topic streaming for intra- and inter-plugin communication (and communication external to the edge server).  

The figure depicts an example pipeline for illustrative purposes.  The \textit{Data Ingest} plugin abstracts the data source (so events now use ZMQ) and may do some cleaning.  The \textit{Data Score} plugin detects shapes and label images before the event is passed to a \textit{DataStore} plugin that chooses to discard, store, or archive images. A \textit{Resource Monitoring} plugin captures resource (CPU, memory, energy) usage and finally, an \textit{Inference} plugin might carry out more refined detection.  The ML Field Planner listens on a topic for results that are directed at the user for results analysis.

 Crucially, the data pushed to Patra includes the original model\_id provided at launch which is used to properly link edge server information to its model card.

\section{System Design - Model Context Protocol\label{sec:design}} 

Model Context Protocol (MCP), recently released from Anthropic, offers a session-based, standardized approach to AI tool interaction that promises improved efficiency for complex, multi-step workflows. MCP standardizes the interaction between AI systems and external tools and resources through a JSON-RPC-based session protocol. 

REST APIs have long served as the backbone of web services communication, providing stateless, scalable interfaces for distributed systems. Patra earlier made the highly reasonable decision to use REST for its API (Patra communication with CKN is topic-based streaming using Kafka). However, the ICICLE ecosystem could benefit from supporting MCP as a single protocol that harmonizes all data resources in the ecosystem under a single protocol. ICICLE components would thus have uniform, discoverable procedures throughout the AI/ML model lifecycle of model selection, performance tracking, provenance gathering, and lifecycle transitions, 
thereby streamlining automation across resource-constrained and heterogeneous settings. Unlike REST's stateless request-response pattern, MCP maintains persistent sessions with capability discovery, enabling multi-operation workflows and contextual tool invocation. Through this session-based communication, ICICLE could enhance data resource interaction during the course of computational research workflows.

MCP declares three core primitives that servers can expose:  resources, tools, and prompts.  As per MCP documentation~\cite{MCP}, \textit{MCP resources} are data sources that provide contextual information to AI applications (e.g., file contents, database records, API responses). \textit{MCP tools} are executable functions that AI applications can invoke to perform actions (e.g., file operations, API calls, database queries). And \textit{MCP Prompts} are reusable templates that help structure interactions with language models (e.g., system prompts, few-shot examples). 

MCP uses the JSON-RPC 2.0 standard over lightweight, text‑based transports such as HTTP(S) or Server‑Sent-Events (SSE) for persistent sessions.

\paragraph{MCP Resources, Tools in Patra} We design Model Context Protocol (MCP) servers that serve the functionality of Patra.   Our implementation extends the FastMCP framework to provide specialized functionality for the Patra Knowledge Graph, incorporating performance monitoring, caching mechanisms, and batch processing capabilities designed specifically for scientific model management scenarios.

We base our performance study on two Patra operations:

\begin{itemize}
     \item \textbf{Model Card Retrieval}: Retrieves the full model card for a given model identifier. 
     \item \textbf{Edge Creation}: Creates an edge (directed relationship) between two nodes in the graph.  The operation validates node existence, infers relationship type from node labels via schema constraints, checks for duplicate edges, and commits the edge creation within an atomic transaction. 
\end{itemize}

\textit{Model card retrieval} is implemented through the MCP "resource" primitive because it provides contextual information about a model card in a read-only manner. Specifically, it implements  the URI pattern \texttt{modelcard://\{mc\_id\}}, where \texttt{\{mc\_id\}} is the model card external identifier. Retrieval employs a multi-query aggregation strategy within a single Neo4j session: a Base query that retrieves the primary ModelCard node, and four independent queries fetch related entities: AI Model, Bias Analysis, Explainability Analysis, and Deployments. The results are aggregated into a nested dictionary and serialized to JSON.

We implement the \textit{edge creation} operation that adds an edge to the graph representing a model card. This operation is implemented through the MCP "tool" primitive. \texttt{create\_edge} creates directed relationships between two nodes, parameterized by Neo4j identifiers. This tool is used in the Patra-RGCN~\cite{patra-rgcn}, an extension of Relational Graph Convolutional Networks for link prediction. Relationship types are automatically inferred from node label pairs using a static mapping. 

The \texttt{create\_edge} MCP tool executes a four-stage validation and commit pipeline to ensure schema compliance and idempotency: 1) The tool accepts two Neo4j \texttt{elementId} parameters and issues a Cypher query to verify both nodes exist and retrieve their label sets. %If either node is missing or unlabeled, the tool returns an error response immediately.
2) The tool normalizes node labels and validates the relationship against a dictionary, which encodes the knowledge graph schema as a directed adjacency structure.
3) Before creating the edge, the tool checks whether a relationship of the inferred type already exists between the two nodes.
4)If all validation passes and no duplicate exists, the tool commits the relationship within a single Neo4j transaction.

\section{Performance Evaluation \label{sec:experim}}

To better understand the benefits of MCP for our uses, we develop two MCP servers for comparison against the existing Patra REST API. We execute end-to-end requests using two popular Patra calls: model card retrieval (get a model card) and model card search. The three server versions are thus:

\begin{itemize}
\item \textit{REST}: Client invokes the Patra REST API directly via HTTP, with each request establishing a stateless connection.

\item \textit{Native MCP}: the MCP server executes queries directly against the Neo4j database using the Bolt protocol, bypassing the REST layer entirely.

\item \textit{Layered MCP}: MCP server interacts through the Patra REST layer, translating calls from the MCP tools to their correpsonding REST endpoints. 
\end{itemize}

Where MCP is involved, both functions are implemented as MCP tools in Python, following the standard \texttt{@mcp.tool} decorator pattern. The \textit{Native MCP} implements the same logic as the REST endpoint does with respect to requiring multiple database queries but then wraps the response in JSON‑RPC 2.0 message framing. \textit{Layered MCP} adds a third layer: the MCP tool internally calls the REST endpoint which introduces a HTTP client‑server handshake and duplicate JSON serialization (once for the REST response, again for the MCP JSON‑RPC envelope). 

\paragraph{Test Environment}
We carry out two experiments: a microbenchmark experiment and a real world experiment. The microbenchmark experiment uses two nodes from the Jetstream2 system \cite{jetstream2}. The client node is a m3.quad instance with 4 CPU, 15 GB RAM, and 20 GB hard drive. The server node is a m3.medium instance with 8 CPU, 30 GB RAM, and 60 GB hard drive. All instances run  Ubuntu 24.04.3 LTS
operating system and are hosted in Indiana USA. The Jetstream2 compute nodes are connected through a 100 Gbps network. The multiple runs of our experiment show no perceptible network interference.

For the real world experiment, the client node is of the same size but resides in Hawaii USA. The server node is of the same size and location in Indiana USA. All instances run Ubuntu 24.04.3 LTS operating system.

The Model Context Protocol (MCP) server and client are implemented using the MCP\,[CLI] library~(v1.10.1). All experiment scripts utilize the MCP Python SDK for establishing Server-Sent Events (SSE) transport sessions. Each benchmark uses asynchronous I/O routines specifically Python’s \texttt{asyncio} module to minimize client-side variability. The graph database is Neo4j v5.21.0.

The server components run as isolated Docker containers (Docker Engine v28.2.2) with explicit CPU and memory limits to reduce noisy-neighbor effects and improve reproducibility. Neo4j is allocated 6 vCPU and 12~GB memory; all API servers run with 4 vCPU and 4~GB memory. The containers communicate over a Docker bridge network.

% \color{red}neelesh, discuss what goes into the VMs that we're building and using for this. For instance, we're using docker so running services as containers.  what version of docker?  what else are we using that we've not called attention to here?

\paragraph{Graph database} For the microbenchmark experiment, we use model cards of a few KB in size. The graph database for the real-world experiment consists of 10,000 Deployment nodes, 1000 Experiment nodes, and 100 Device nodes. This is pseudo-synthetic data. Based on real model cards, we extended the deployment, experiment, and device information.  For the wide area experiment, a single model card including its deployment information averages 13.63~MB (measured in its JSON format.)

\subsection{Experiment I: Microbenchmarks}
The first experiment we undertake is a microbenchmark to capture both end-to-end and component latency. The layered MCP interface wraps existing REST endpoints through JSON-RPC translation, the native MCP implementation invokes data access logic directly through the FastMCP framework. Each benchmarked operation executes over a persistent SSE session and is repeated 1,000 times for statistical consistency.  The experiment API endpoints return small payloads, on the order of KBs. 
%These operations emphasize JSON-RPC serialization and request routing latency impacts, making them ideal for quantifying protocol-layer overhead.  

The total query time for retrieving (few KB) model card is $5.8\pm0.7$ ms for model card retrieval 
%and $1.3\pm0.5$ for \texttt{search\_modelcards} 
operation.  Model card retrieval is actually several queries, as shown in Table~\ref{tab:database_retrieval1}. This represents the theoretical lower bound for query execution time

\begin{table}[tbh]
\centering
\caption{Database Retrieval Time}
\label{tab:database_retrieval1}
\begin{tabular}{l r}
\hline
\textbf{Node} & \textbf{Latency (ms)}\\
\hline
Model Card & 1.19\\
Model & 1.0\\
Bias Analysis & 1.04\\
XAI Analysis & 0.95\\
Deployments & 1.68\\
\hline
Total & 5.87\\
\hline
\end{tabular}
\end{table}

\begin{figure}[tbh]
    \centering
    \includegraphics[width=8.0cm]{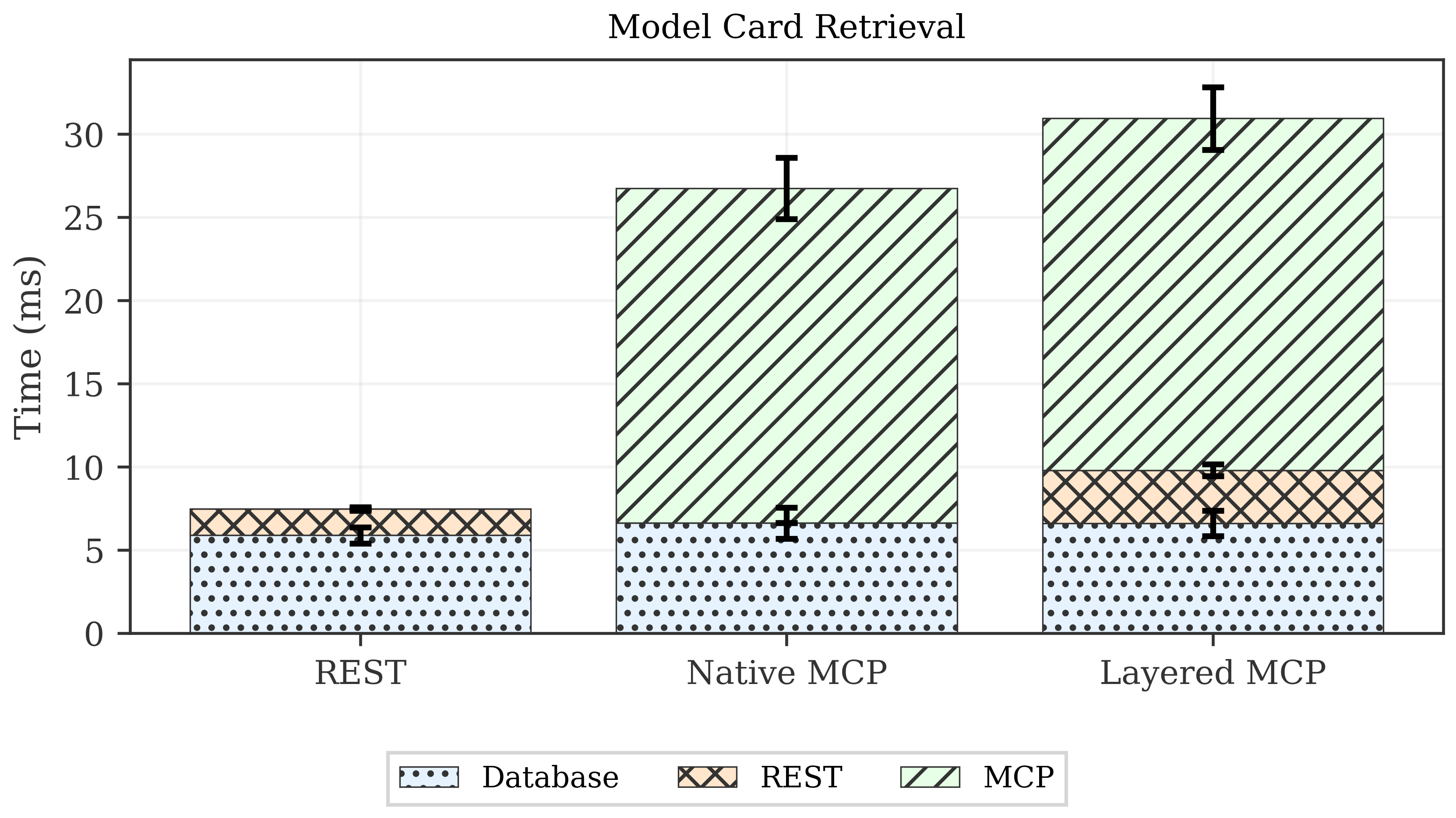}
    \caption{Model card retrieval, few KB model card sizes}
    \label{fig:local_modelcard}
\end{figure}

In comparing across the three server variants for the model card retrieval operation (for KB model cards), as shown in Figure ~\ref{fig:local_modelcard}, REST has the most efficient protocol implementation, with 7.5 ms total retrieval time.  Its small deviation reflects Flask's mature HTTP handling and direct JSON serialization paths.

The Native MCP server has a 3.6 times higher turnaround time than the REST server does for model card retrieval with 26.7 ms. This performance differential stems from MCP’s JSON-RPC 2.0 protocol complexity, including message envelope construction, tool dispatch mechanics, and SSE session management~\cite{jackson2024streaming}. 
The layered REST+MCP architecture is 4.1 times the turnaround as does vanilla REST; this exposes a  double serialization penalty: the initial Flask JSON encoding followed by MCP JSON-RPC envelope wrapping. The additional REST layer for the REST+MPC is a 16\% overhead from native MCP.  While operationally convenient for maintaining backward compatibility, the layered architecture’s 4× overhead relative to REST establishes it as suitable primarily for development environments and transitional deployments rather than production systems optimizing for performance.

The difference between native and layered MCP architectures centers on protocol integration depth and data flow abstraction. In a native implementation, the Model Context Protocol (MCP) interacts directly with the application layer, bypassing any intermediary HTTP-based abstraction such as REST. This design aligns with direct binding theory where minimal protocol layering leads to reduced serialization overhead and tighter coupling between semantic logic and transport mechanisms~\cite{jackson2024streaming}. Native MCP servers perform all operations within the MCP tool abstraction, handling JSON-RPC message parsing, state management, and response serialization directly. The resulting simplicity in data flow can reduce latency but requires re-implementation of existing REST logic and error-handling layers, increasing development complexity. In contrast, the layered MCP architecture applies an adapter pattern where MCP acts as a proxy wrapping existing REST endpoints. This approach leverages REST’s mature API contracts and established tooling while exposing equivalent functionality through MCP’s tool and resource schemas.

Layering preserves backward compatibility and accelerates adoption but introduces double serialization costs and compounded transport overhead. The choice between the two approaches, therefore, reflects a tradeoff between architectural purity and integration pragmatism—native designs favor performance and stability within agentic infrastructures, while layered designs favor interoperability and evolutionary system transitions.

 In wide-area scenarios, connection time escalates, suggesting that SSE connection setup involves multiple network round-trips. The server-side Neo4j execution time is identical to the REST implementation. In Native MCP, database latency is comparable to that of REST, validating that protocol overhead does not impact backend query performance. Server Processing latency includes MCP session management, response framing (SSE encoding), JSON serialization, and accumulated delays from the persistent connection model. In a Layered MCP implementation, the REST overhead is attributed to the additional cost of marshaling MCP operations into HTTP requests and sending them to the REST API. For edge creation, REST overhead is constant, representing HTTP serialization and deserialization, as well as network transmission to the REST layer. For model card retrieval, REST overhead becomes substantial, indicating that transmitting large payloads through the REST layer incurs measurable latency beyond direct database access.

\subsection{Experiment II: Real World}

 In this second experiment we drive towards a more realistic scenario and do so in a few dimensions.  As indicated earlier, Patra is unique in that it accumulates usage information about model inference servers. As such, model cards can grow to be sizeable.  In this experiment we use synthetic model cards that are 13 MB in size (when formated as a JSON object).  

 We further create a second client that is located on Jetstream2 servers located in Hawaii (with the Patra server still residing in Indiana).  We evaluate the two operations using substantially larger model cards and introducing a wide area dimension to the testing. 
 
% The round-trip time using averaged \textbf{97.853 ms} (min: 97.723 ms, max: 98.380 ms, mdev: 0.107 ms) when the server was in Indiana and client was in Hawaii.

With large model cards (13.63 MB), the database query response time grows to 7843 ms, see Table~\ref{fig:dblatency}. This compared to the retrieval time of a small (few KB) model card which is under 6 ms.

\begin{table}[tbh]
\centering
\caption{Model Card Retrieval; large model card}
\label{fig:dblatency}
\begin{tabular}{l r}
\hline
\textbf{Node} & \textbf{Latency (ms)}\\
\hline
Model Card & 7843.83\\
\hline
\end{tabular}
\end{table}

In measuring client-server end-to-end latency, timing is broken down into the following components:

\begin{itemize}
\item{\textit{Connection Setup}}: captures the time required to establish a 
%Server-Sent Events (SSE) 
connection between the client and the server. It includes the TCP handshake and the creation of a persistent HTTP connection for event streaming. Connection setup is a one-time cost and is analogous to the initial connection establishment in traditional HTTP protocols. 

\item{\textit{SSE Handshake}} Server-Sent Events (SSE) is a transport session connection between the MCP client and server. It involves socket initialization and HTTP upgrade negotiation. Local connection overhead includes the initialization of the MCP session. MCP experiment scripts utilize the MCP Python SDK for establishing Server-Sent Events (SSE) transport sessions.  SSE is used in MCP but not REST to maintain the session between requests (REST is stateless).

\item{\textit{Server Processing}}: the time taken to issue a resource read command over the established session and receive the response. It includes network transfer, protocol overhead, server-side processing, and backend database access culminating in the delivery of the requested resource back to the client.
\end{itemize}

\begin{figure}[tbh]
    \centering
    \includegraphics[width=8.0cm]{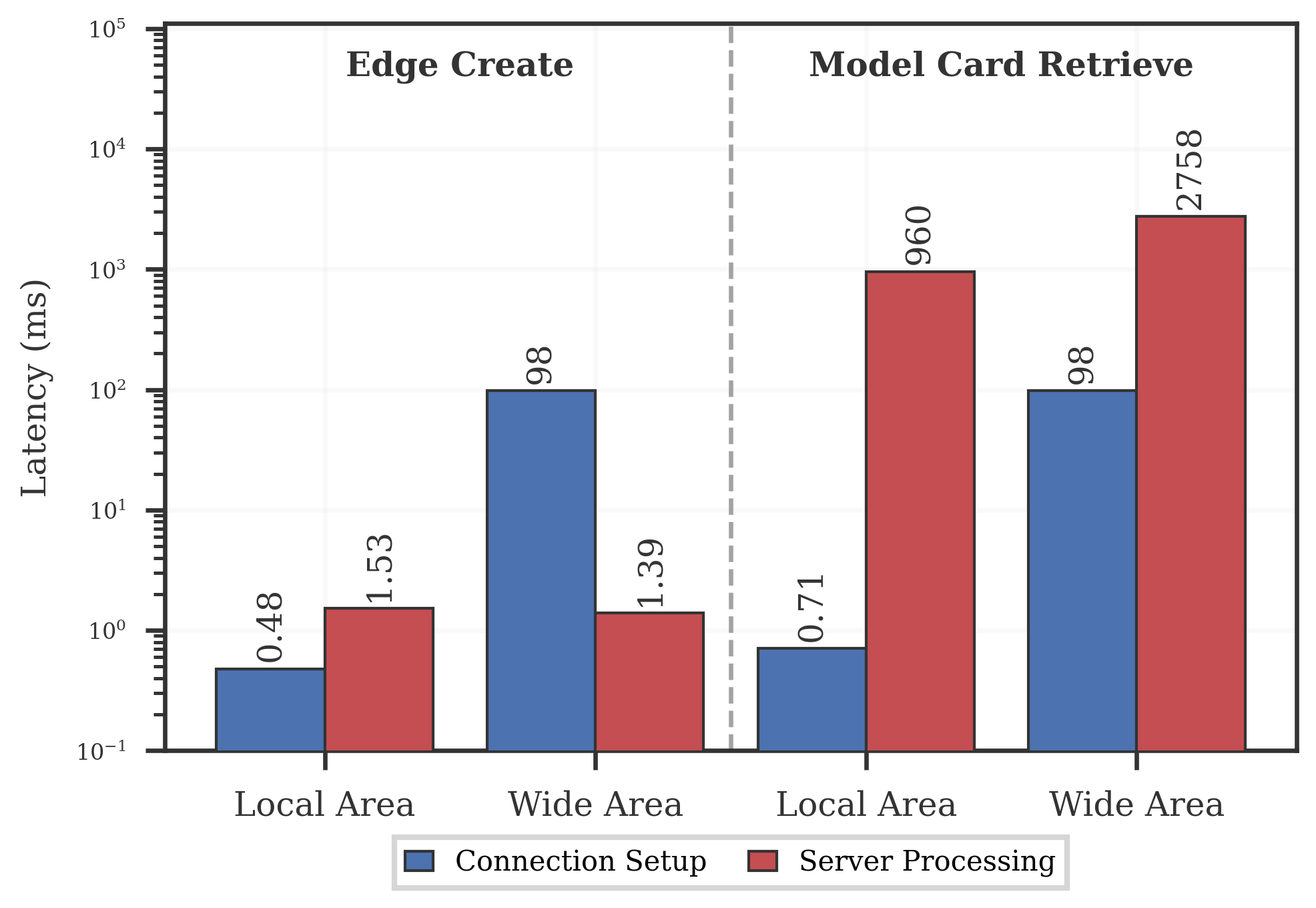}
    \caption{REST, large model card}
    \label{fig:REST}
\end{figure}

\begin{figure}[tbh]
    \centering
    \includegraphics[width=8.0cm]{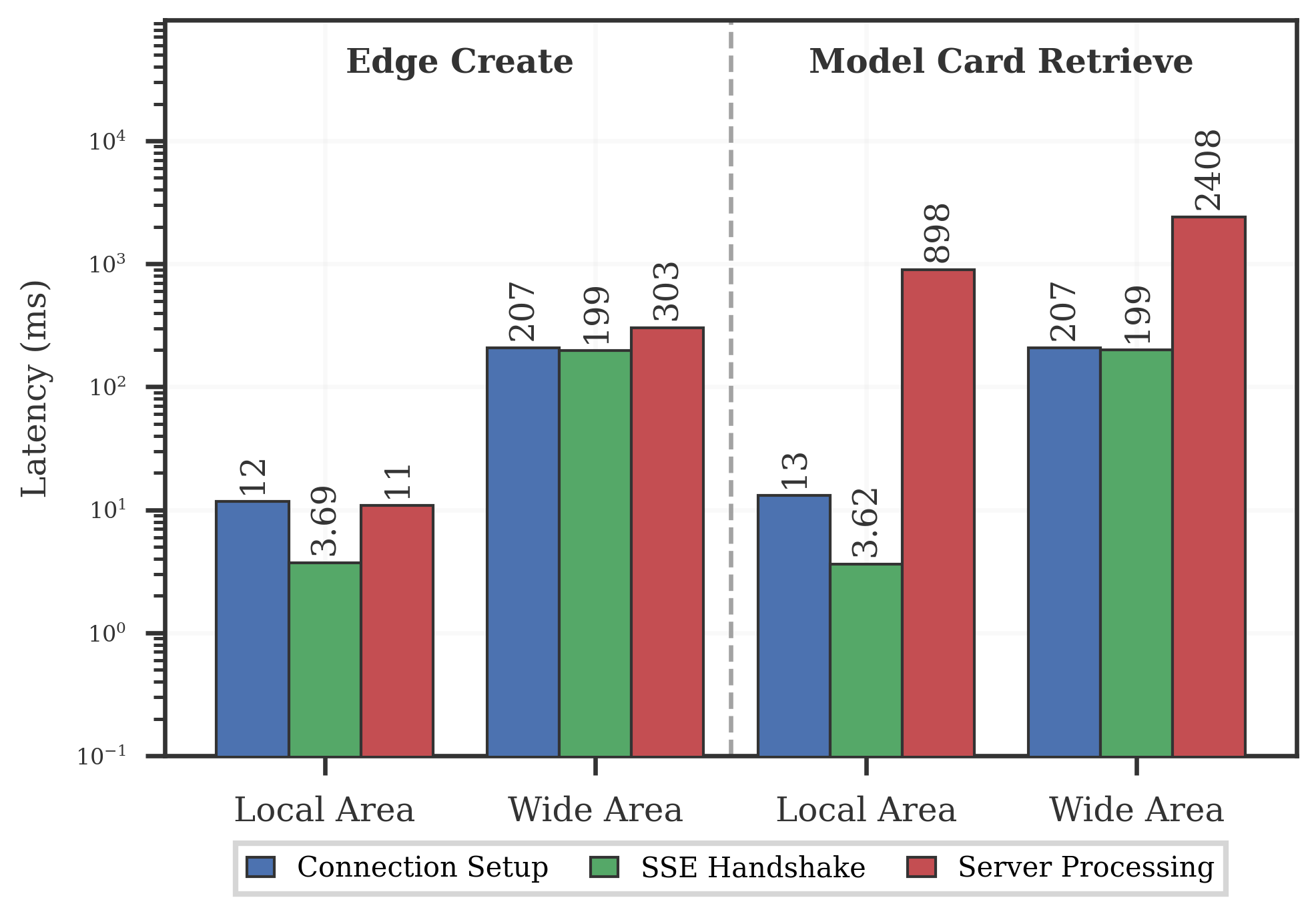}
    \caption{Native MCP, large model card}
    \label{fig:native MCP}
\end{figure}

\begin{figure}[tbh]
    \centering
    \includegraphics[width=8.0cm]{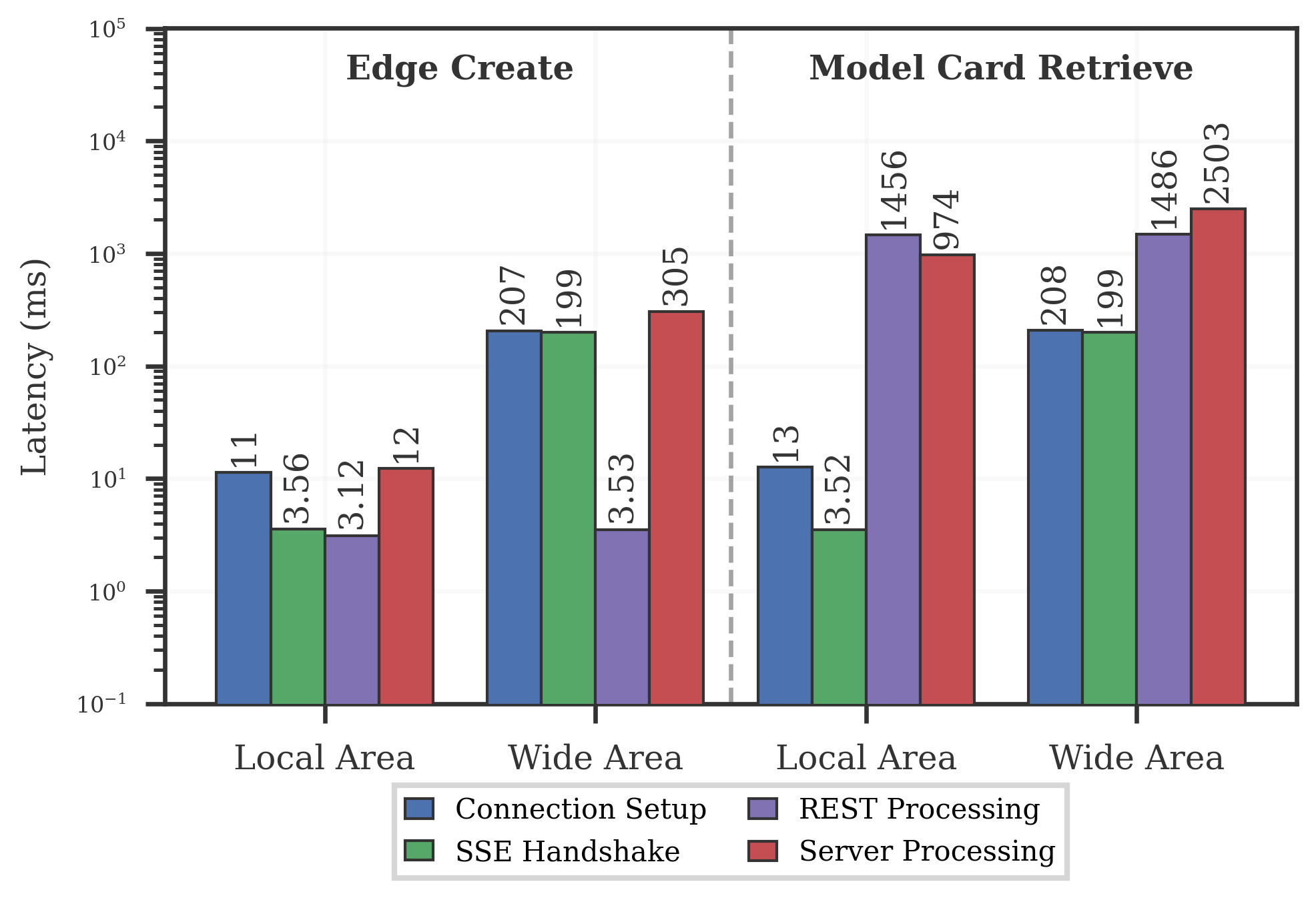}
    \caption{Layered MCP, large model card}
    \label{fig:layered MCP}
\end{figure}

The REST results (shown in Figure~\ref{fig:REST}, show connection setup times that are fairly independent of the distance between client and server.  As server processing includes data transfer back to the client, the wide area numbers reflect the large payload being returned.  Note the log scale used in all of Figures ~\ref{fig:REST},~\ref{fig:native MCP} and ~\ref{fig:layered MCP}.

In comparing native MCP to layered MCP for Model Card Retrieval and the large model card (Figures~\ref{fig:native MCP} and ~\ref{fig:layered MCP} respectively), one can observe that with REST set aside in Figure~\ref{fig:layered MCP} the latencies track one another pretty closely. What this suggests is that the use of REST does not particularly penalize MCP.  It does, however introduce its own overheads that are evident in purple in Figure~\ref{fig:layered MCP}.  In comparing REST (Figure~\ref{fig:REST} and native MCP~\ref{fig:native MCP}, one can see sizeable overheads in the wide area setting for Edge Create. Edge Create returns few KB result sets, so there is additional protocol overheads with MCP in all of connection setup, SSE Handshake, and Server processing.  MCP is 7 times slower than REST for this particular type of operation.

%where the MCP server delegates resource access to a REST backend, SSE connection (mean: 222.13~ms), handshake (mean: 199.46~ms), and resource read (mean: 231.35~ms), but with higher total latency (mean: \textbf{652.94~ms}) due to the intermediate REST API hop (mean: 30.21~ms) and accumulated serialization/deserialization cost. In this stack, database retrieval doesn't change (mean: 16.27~ms, median: 15.14~ms). Together, these measurements highlight that while network setup and protocol overhead are the dominant contributors to end-to-end latency in wide-area deployments, both MCP variants maintain a clear demarcation between baseline data access and the cumulative cost of protocol orchestration, especially when layered abstractions are present.\color{red}neelesh, similarly, make the numbers in the narrative match table II exactly. They're slightly off\color{black}

\begin{figure}[tbh]
    \centering
    \includegraphics[width=8.0cm]{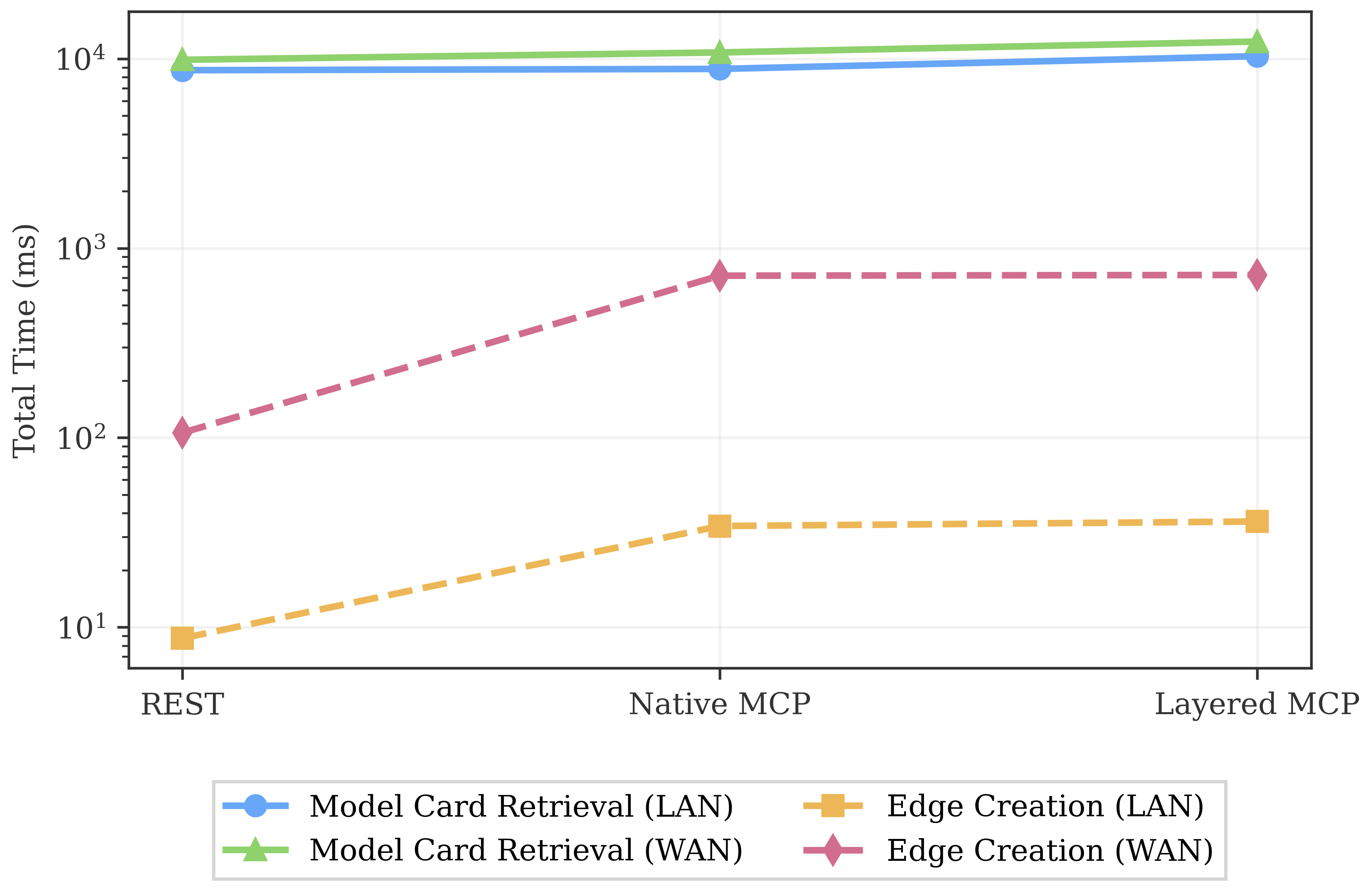}
    \caption{Total turnaround time for each approach on a log scale.}
    \label{fig:total turnaround}
\end{figure}

We conclude in Figure~\ref{fig:total turnaround} by plotting total end-to-end latencies for the local versus wide area configurations and the three server versions (REST, native MCP, and Layered MCP).  Note the log scale on the Y axis.  What this shows is that when result sets are large, the difference between Native MCP and Layered MCP, and even including REST, are not that large. This holds even more true when wide areas are involved.

\section{Discussion}\label{sec:disc}
%We establish quantitative baselines for REST, native MCP, and layered MCP implementations. Our findings show that REST maintains the lowest overhead, confirming mature HTTP stacks' efficiency for lightweight data queries. 

One of the most interesting outcomes of the quantitative evaluation is the additive nature of REST + MCP. That is, when one layers REST + MCP, one pretty well takes on the overheads of both protocols. But in the broader setting of big result sets, non-trivial database graph walks, and longer distances, the significance of the additional overhead diminishes in importance.

A fundamental decision point with MCP adoption boils down to whether a data-holding service can take advantage of MCP's session orientation which is supported through the Server Side Events (SSE) protocol.  SSE exists to support server-side notifications, which give a server the mechanism to push out notifications/events over time to a listening client, often an AI Agent.  

Notifications imply the existence of a client that wants to know about change over time in something.  Suppose, for example, an AI Agent maintains a internal model of the usage of a particular AI/ML model or set of models over time. Using the historical context that an LLM could process, the agent could flag usage that deviates from historical use. The role of the Patra repository then is to notify the agent each time there is a new deployment of a model card. 
   
More broadly this interaction between data store and AI agent is cast as an interaction method such as the ReAct (Reson + Act) method ~\cite{masterman2024landscapeemergingaiagent}.  In this method, an agent presents the state of the world to an LLM, solicits LLM response, then asks the LLM what it wants to do next. An agent begins by writing a thought about the given task. The agent then performs
an action based on that thought, and the output is observed. This cycle can be repeated until the task is complete. 

As Masterman \textit{et al.}~\cite{masterman2024landscapeemergingaiagent} point out, however, the ReAct method is not without its limitations. "While intertwining reasoning, observation, and action
improves trustworthiness", they state, "the model can repetitively generate the same thoughts and actions and fail to create new
thoughts to provoke finishing the task and exiting the ReAct loop." 

Circumventing the repetitive loop of the same thoughts and actions from the LLM agent could be accomplished by incorporating human feedback during the execution
of the task.  

%When viewed from the lense of protocols like MCP, human-agent interaction could be facilitated through the protocol by the transfer of the current session between agent to human, agent to agent, and agent to a persistent storage all conceivably through MCP. This session state could additionally be cached at MCP servers, and persisted in MCP-enabled storage systems.  

We turn back to the initial argument in the paper:  what does it mean to benchmark model use at points during its lifecycle rather than just at training time.   We interpret benchmarking model use to having a "by whom" component and a "how" component.  

%With the introduction of LLMs, context has become critical. The absence of state in REST could in the end undermine use of this popular protocol. 
%It is an open question of whether or not future uses of a model card repository are better served through a session interface. That is to say, are inquiries to the repository better served by retention of prior requests against the database, such as might be the case for graph traversal needed for missing edge detection. Or should Patra find itself as a resource in agent to agent chaining, the context could be important.  
%Future work explores this question. 
%Additionally, MCP's session persistence may prove advantageous when round-trip time dominates protocol overhead, potentially offsetting JSON-RPC serialization costs in high-latency environments.
%The current benchmarks focus on queries with small result sets. In large payload transfers, MCP's persistent connections could amortize handshake overhead across multi-gigabyte transfers, while REST's stateless design faces repeated connection establishment costs. The experiments were executed within a single Jetstream2 availability zone with no perceptible network interference. Benchmarks across geographically distributed sites would quantify protocol overhead under realistic network conditions.

 To the former, "by whom", we are cognizant that model cards belong to the person who created the model. There are many implications of model card ownership that we do not go into here, but before describing how we track the creator, we point out that an individual's organizational affiliation is a strong indication of how a model has been used.  If a model is used by a county courthouse, this tells us something about its use.  Organizational affiliation is a placeholder for us - an indicator of use that we intend to push on more deeply. 
 
 Model Cards have the same creator as the AI/ML model by the following: using the toolkit’s submit() method, the user can simultaneously upload the model artifacts to a selected repository (e.g., HuggingFace or GitHub) and ingest the Model Card to the Patra server.  Ownership is established through the Tapis hosted infrastructure API.  Tapis provides the authorization and authentication framework for the ICICLE cyberinfrastructure, maintaining the identity of individual researchers using the platform.  A researcher can, for instance, retrieve a trained model artifact using their credentials or share it with a collaborator via a time-limited access token. The integration ensures that an actual user of the ICICLE environment is associated with each model card in the system.

As to the "how", as captured in the data model of Figure~\ref{fig:dataModel}, our group is exploring the capture and representation of information surrounding the use of a model in the form of the experiment (or task graph in agentic AI).  For the model card to have the needed information to deploy the right model we think will give us leverage, along with the "by whom" to trigger alerts when a model is being used outside of some tolerance of its deployment in the past.  This is future work for us.

\section{Related Work\label{sec:related}}
% \color{red}neelesh, please work this section. no footnotes (cite published sources whenever possible and create citation for URL when must).  it's generally in pretty good shape.  Other work on model cards (citable from published literature) are helpful \color{black}

The Patra Knowledge Base runs as a persistent service, ingesting JSON model cards as graph representations using, in part, the PROV-ML ontology~\cite{belhajjame2013prov}. Several recent projects in AI MLOps and distributed systems have begun to adopt agent-based architectures, but only a few leverage open protocols for tool/resource discovery. MCP advances the state of the art by formalizing the way agents communicate, interrogate, and control external resources (such as artifact registries, policy engines, and provenance graphs) in a vendor- and language-neutral manner. Within Patra and CKN, this means that model lifecycle, deployment, and provenance steps may be accessed and controlled by any agent that speaks MCP, simplifying integration with major AI ecosystems or federated scientific infrastructures. 

Protocol performance evaluation has been extensively studied in the context of web services. Pautasso et al. \cite{pautasso2008restful} established foundational principles for REST API design and performance characteristics. Recent work by Zhang et al. \cite{zhang2023grpc} compared gRPC, REST, and GraphQL across e-commerce workloads, finding significant performance variations based on operation complexity and data access patterns. However, existing studies focus primarily on traditional web service scenarios. The unique characteristics of AI model management—including complex metadata structures, multi-criteria decision making, and edge deployment constraints—have not been systematically evaluated across different protocol architectures. 

%The Model Context Protocol represents a recent development in AI system interoperability. Our work fills this gap by providing an empirical evaluation of both protocols under controlled conditions.

Comparable efforts in reproducibility and provenance-aware scientific workflows include the Sciunit framework by Tanu Malik et al. \cite{malik2017sciunits}, which encapsulates scientific processes as reusable research objects with captured provenance for transparent re-execution. Recent works leverage large language models to automate documentation using pipelines like CardGen \cite{liu-etal-2024-automatic}. 
% Early work on optimization of inference at the edge uses compressed deep networks with weight pruning and low‑bit quantization. Beyond single‑model compression, recent work considers the holistic selection of models under accuracy--latency--cost trade‑offs. Zhou et al~\cite{Zhou2025} formulate multi‑model pull decisions for edge cloudlets as an Integer Linear Program, sharing parameters across fine‑tuned models to lower total delay and cost by at least 38\% and improve accuracies by 5\%.
% MLOps platforms support AI/ML pipelines through creation, deployment, and continuous delivery. These platforms, such as MLFlow \cite{zaharia2018accelerating}, provide tools for experiment tracking, model packaging, and versioning. Rani et al. \cite{RANI20243019} underscore the need for robust MLOps pipelines that can span cloud and edge. 
% Lorenzo et al. \cite{colombi2024mlops} shows how a unified MLOps solution can enable continuous delivery of models in an edge-cloud environment. The study demonstrates effective deployment of models to an edge gateway and cloud backend with a unified interface.

LinkEdge \cite{dias2023linkedge} is a lightweight platform that bridges cloud MLOps pipelines with edge devices. The open, modular nature of LinkEdge aligns with Patra's goals of scalable and transparent edge-cloud ML management. In our architecture, Patra issues globally unique PIDs for every serialized model and version of its ModelCard, while
TAPIS \cite{tapis} is the underlying object store and resolution fabric.
Patra Model Cards employ signposting. The seminal work of Van de Sompel et al. \cite{fairsignposting} introduces FAIR Signposting Profile as implementation guidelines for exposing machine-actionable navigation links using standardized HTTP headers and HTML link elements. Soiland-Reyes et al. \cite{soiland2025rocrate} demonstrates the combination of signposting with RO-Crate as a practical, web-centric approach to implement FAIR Digital Objects. %We adopt signposting to advertise the location of a Patra ModelCard, its associated PID record, and runtime artifacts such as provenance logs, enabling automated harvesting by downstream governance tools.
Kreuzberger et al.~\cite{kreuzberger2023mlops} point out that MLOps platforms often rely on internal identifiers. Internal identifiers are inconsistent with the principles of FAIR software and data and open science.  

%athikar et al. \cite{Samtani2023} access the security vulnerabilities of hosting models in HuggingFace and underscores the importance of having information available to make informed decisions about model use. 
Renan et al. \cite{santos2024workflow} articulates the vision for using workflow provenance across the edge-cloud continuum to support Responsible and Trustworthy AI. They argue that provenance data is critical but note a gap in current systems where provenance management is not well-integrated with heterogeneous infrastructures and ML life-cycles. Venkataramanan et al. \cite{venkataramanan2023kg} demonstrate the scalability of using Knowledge Graphs to integrate diverse ML metadata from multiple sources for querying and recommendation.

% Wang et al. \cite{wang2023preventing} addresses the challenge of maintaining algorithmic fairness in an online machine learning setting where data is streaming and evolving. They propose a technique called FS² (Fair Sampling over Stream), which dynamically rebalances the input stream to mitigate bias while accounting for concept drift. They also introduce Fairness-Bounded Utility (FBU), a unified metric to evaluate the trade-off between accuracy and fairness for streaming models. Khalyeyev et al. \cite{Khalyeyev_2023} reviews the existing notions of the edge-cloud continuum and frames the broader vision in which CKN-Patra operates. 

\section{Conclusion and Future Work}\label{sec:concl}

In this article we contextualize model cards within the larger setting of the ICICLE AI Institute software ecosystem.  We quantify overheads of REST versus Model Context Protocol (MCP) protocols in serving model cards in various settings.  We call out the most significant distinction between REST and MCP, and that is the session orientation of the latter.  This we explore in terms of agentic AI and whether Patra model cards' goal of AI accountability can be reached by its support of benchmarking model use at points in model lifecycle (beyond its initial training).  We are encouraged.

An additional open question addresses Patra's 
%Scalability testing under concurrent load would reveal whether MCP's session management introduces server-side bottlenecks (connection pooling, message queue depth) or whether persistent connections reduce per-client overhead at scale. This analysis should incorporate resource utilization metrics to identify scaling limits.
heavy dependency on attributes/constraints attached to nodes and tasking of REST endpoints with sole knowledge of schema information (such as what defines a model card). These are two major design choices that need further study in light of AI interactions with the repository.

\section{Acknowledgements}\label{ack}
We thank the entire ICICLE Smart Fields team, but particularly Christopher Stewart of Ohio State University. %This research is funded in part through the National Science Foundation NSF ICICLE AI Institute award \#2112606 and through the Data To Insight Center at Indiana University. We thank the reviewers for thoughtful feedback.

\bibliography{bibliography}
\bibliographystyle{plain}

\vspace{12pt}

\end{document}